\def\papertitle{Sampling-rate-aware noise generation}
\def\paperauthorA{Henning Thielemann}

\documentclass[twoside,a4paper]{article}
\usepackage{dafx_11}
\usepackage{amsmath,amssymb,amsfonts,amsthm}
\usepackage{euscript}
\usepackage[latin1]{inputenc}
\usepackage[T1]{fontenc}
\usepackage{ifpdf}

\usepackage[english]{babel}
\usepackage{caption}
\usepackage{subfig, color}

\ninept

\usepackage{times}
\ifpdf % compiling with pdflatex
  \usepackage[pdftex,
    pdftitle={\papertitle},
    pdfauthor={\paperauthorA},
    colorlinks=false, % links are activated as color boxes instead of color text
    bookmarksnumbered, % use section numbers with bookmarks
    pdfstartview=XYZ % start with zoom=100% instead of full screen; especially useful if working with a big screen :-)
  ]{hyperref}
  \usepackage[pdftex]{graphicx}
  \usepackage[figure,table]{hypcap}
\else % compiling with latex
  \usepackage[dvips]{epsfig,graphicx}
  \usepackage[dvips,
    colorlinks=false, % no color links
    bookmarksnumbered, % use section numbers with bookmarks
    pdfstartview=XYZ % start with zoom=100% instead of full screen
  ]{hyperref}
  \usepackage[figure,table]{hypcap}
\fi

\title{\papertitle}

\usepackage{tikz}
\usetikzlibrary{arrows}

\usepackage{comment}
\usepackage{verbatim}
\usepackage{mathpaper}
\usepackage{IEEEtrantools}

\theoremstyle{definition}
\newtheorem{criterion}[definition]{Criterion}
\newcommand\crtlabel[1]{\label{crt:#1}}
\newcommand\crtref[1]{Criterion~\ref{crt:#1}}

\DeclareMathOperator{\NSD}{NSD}
\DeclareMathOperator{\VSD}{VSD}
\DeclareMathOperator{\IDFT}{DFT^{-1}}
\DeclareMathOperator*{\render}{render}
\DeclareMathOperator*{\resample}{resample}
\DeclareMathOperator*{\filter}{filter}
\DeclareMathOperator*{\noise}{noise}
\DeclareMathOperator*{\spectrum}{spectrum}
\DeclareMathOperator*{\clip}{clip}
\DeclareMathOperator*{\weight}{weight}
\DeclareMathOperator*{\quantise}{quantise}
\newcommand\Var{\sigma^2}
\newcommand\stddev{\sigma}
\newcommand\expectedvalue{\mathbb{E}}

\newcommand\reviewernote[1]{}

\newcommand\code[1]{\texttt{#1}}

\newcommand\unit[1]{\text{#1}}
\newcommand\hz[1]{#1~\unit{Hz}}

\newcommand\figcaption[2]{\caption{\textit{#2}}\figlabel{#1}}

\affiliation{
\paperauthorA}
{\href{http://www.informatik.uni-halle.de}{Institut f\"{u}r Informatik,
Martin-Luther-Universit\"{a}t Halle-Wittenberg, Germany} \\
{\tt \href{mailto:henning.thielemann@informatik.uni-halle.de}{henning.thielemann@informatik.uni-halle.de}}
}
\begin{document}
% more pdf-tex settings:
\ifpdf % used graphic file format for pdflatex
  \DeclareGraphicsExtensions{.png,.jpg,.pdf}
\else  % used graphic file format for latex
  \DeclareGraphicsExtensions{.eps}
\fi

\maketitle

\begin{abstract}
In this paper we consider the generation of discrete white noise.
% In this paper we consider the generation of discrete white noise,
% i.e.\ an observation of a sequence
% of independent and identically distributed random variables with expectation value 0.
Despite this seems to be a simple problem,
common noise generator implementations
do not deliver comparable results at different sampling rates.
First we define what we mean with ``comparable results''.
From this we conclude, that the variance of the random variables
shall grow proportionally to the sampling rate.
Eventually we consider how noise behaves under common signal transformations,
such as frequency filters, quantisation and impulse generation
and we explore how these signal transformations must be designed
in order generate sampling-rate-aware results when applied to white noise.
% We propose that software synthesizers
% shall use the noise spectral density or related parameters
% for specifying white noise.
% \reviewernote{
% For the presentation I plan to show
% some examples of generating sounds that contain noise.
% }
\reviewernote{
The organisation of this article follows the advises of
% \cite{peyton-jones2004paper}.
\href%
{http://research.microsoft.com/en-us/um/people/simonpj/papers/giving-a-talk/giving-a-talk.htm}%
{http://research.microsoft.com/en-us/um/people/simonpj/papers/giving-a-talk/giving-a-talk.htm}.
}
\end{abstract}

\section{Introduction}
\seclabel{introduction}

Noise is an ubiquitous kind of signal:
Often it is an annoying artefact of signal transmission, conversion, or processing,
but it is also an essential part of sounds
like wind, breaking water waves, wind instruments,
drum sounds, and fricatives in speech.
In our paper we explore the generation of the latter kind of noise.
% The precise problem is to generate discrete noise signals in a way
% that they behave comparable for different sampling rates.

\subsection{A motivating example}

Imagine a sound designer who works on a collection of synthesised instruments
that shall be used in software synthesizers.
For instance he tries to match the sound of a panpipe
by mixing a sine oscillator and white noise,
that is filtered by a resonant low-pass filter
as illustrated in \figref{panpipe-flow}.
Then the sound designer goes to implement the simple flow diagram
in the software synthesis package Csound
as shown in \figref{panpipe-csound}.
Since he intends to use the instruments in music for compact discs,
he chooses to render his sound at 16 bit resolution and \hz{44100}.
He achieves this with the following command line.
\begin{verbatim}
csound -o panpipe.wav \
   panpipe.orc panpipe.sco \
   --sample-rate=44100 --control-rate=441
\end{verbatim}
Our artist will later add an envelope, frequency modulation
and other enhancements that make the sound more natural.
He will also design several more instruments.

After his collection of instruments has grown to a considerable size
he decides to also prepare preview sounds
at a lower sampling rate for his web site.
To this end he starts Csound with the option \verb|--sample-rate=11025|.
To his surprise some instruments sound quite different at the lower rate.
He expected a worse quality,
but he assumed that instruments would sound essentially the same.
He goes back to the simple panpipe prototype algorithm
and finds out that the noise portion of the sound is considerably louder
at the low sampling rate than at the high sampling rate.
Thus the carefully chosen mixing ratio of noise and sine wave
is lost at the low sampling rate.

Before filing a Csound bug
the sound designer wants to find out, what is precisely the problem.
He checks that the white noise has the same amplitude
at the low and the high sampling rate.
The same is true for the sine wave.
He replaces the noise generator by an oscillator
and observes that filtering a tone
yields the same result at both sampling rates.
That is, it seems to be the combination of noise and filtering
that introduces the unintended volume dependency on the sampling rate.
This is very strange.

\tikzstyle{process}=[inner sep=1.5ex,rectangle,draw=black!50,fill=black!3,thick]
\begin{figure}
\begin{center}
\begin{tikzpicture}[node distance=13ex]
  \node[]            (output)      {};
  \node[circle,draw] (mix)   [left of=output]      {$+$};
  \node[process] (oscillator) [left of=mix]         {oscillator};
  \node[process] (filter)     at ([yshift=10ex] oscillator) {low-pass};
%  \node[process] (filter)     [above of=oscillator] {\mbox{resonant\\low-pass}};
  \node[process] (noise)      [left of=filter]      {noise};
\pgfsetarrowsend{latex}
  \draw (noise.east)      -- (filter.west);
  \draw (filter.east)     -| (mix.north);
  \draw (oscillator.east) -- (mix.west);
  \draw (mix.east)        -- (output);
\end{tikzpicture}
\end{center}
\figcaption{panpipe-flow}{%
Flow diagram (abstract signal processing algorithm)
for a very simple sound
that resembles a panpipe.
}
\end{figure}
\begin{figure}
Orchestra file \path{panpipe.orc}
\begin{verbatim}
nchnls = 1

instr 1
iamp = 30000
anoise   noise    0.3*iamp, 0
acolored lowpass2 anoise, 440, 10
aosci    oscils   0.5*iamp, 440, 0
out      aosci+acolored
endin
\end{verbatim}
Score file \path{panpipe.sco}
\begin{verbatim}
i 1 0 2
e
\end{verbatim}
\figcaption{panpipe-csound}{
The flow diagram in \figref{panpipe-flow}
translated to a Csound program.
The line containing \code{lowpass2} applies a low-pass filter
with resonance at \hz{440} to the previously generated white noise.
The opcode \code{oscils} generates a sine wave also with frequency \hz{440}.
The score file specifies that our instrument with number~1
starts at second zero and stops after two seconds of playing.
}
\end{figure}

Now he becomes curious whether that problem
also occurs in other sound synthesis systems.
He translates his Csound algorithm
to the real-time synthesizers SuperCollider and ChucK
and confirms that they behave exactly the same way.
Finally he gives up and decides to just downsample the sounds
he rendered at \hz{44100}.
The downsampled wave files actually sound fine.

However our artist is still uncomfortable with the observation
that his signal algorithms depend in a non-obvious manner
on the sampling rate.
So far he thought, that his algorithms abstract from the sampling rate.
He had seen his algorithms as an analogy to scalable vector graphic formats,
such as PostScript, PDF or SVG that can be rendered
at any device at any resolution while achieving the maximum possible quality.
He even expected that he could use the same algorithm
both for discrete and analogue signal processing.
% To draw the analogy even further,
% PCM sound formats like RIFF-WAVE or AIFF
% correspond to raster image formats such as PNG or PGM.
% % JPEG's file interchange format (JFIF).
What is the point of sound synthesis compared to sound sampling
if not flexible adaptation to changing sound parameters
and also to the sampling rate?

He is not quite satisfied with downsampling sounds
from \hz{44100} to \hz{11025}
in order to get sampled sounds at \hz{11025}.
This indirection means that he must invest four times of the computation time
of rendering immediately at \hz{11025}
plus time for resampling.
When rendering music directly at \hz{11025}
he could generate 4 times as many channels for spatial effects
or 4 times of the polyphony
of music rendered via the indirection through \hz{44100}.
Vice versa: How can he produce sampled sounds
with maximum possible quality at \hz{96000}
from his algorithms that he designed for \hz{44100} sampling rate?

\subsection{Basic considerations}

The noise generators found in Csound and other packages
do what certainly everyone would do in order to produce white noise:
They run a standard pseudo-random number generator
in order to fill an array with random values from the interval $[-1,1]$
according to a uniform distribution.
\begin{comment}
However for different media we need different sampling rates,
say \hz{8000} as used in (old) telephone transmission,
\hz{11025} for space saving low-quality MPEG streams,
\hz{44100} for CD quality or
\hz{96000} for DVD-Audio.
\end{comment}
Now let us see, what this actually means when this is performed
at different sampling rates.

In \figref{noise-filter} we have white noise of the same duration
both at \hz{11025} sampling rate
($\render(\hz{11025}, \noise)$ in the top-left-corner)
and at \hz{44100} sampling rate
($\render(\hz{44100}, \noise)$ in the top-right-corner).
The noise at the higher sampling rate looks more dense
than that at lower rate, of course.
We also hear clearly the additional high frequencies in the high rate noise.
However we have the impression,
that the low frequencies of the noise
are louder in the low rate noise
and softer in the high rate noise.
This auditory impression becomes even visual
if we apply frequency filters to this noise.
We have applied a first order low-pass filter
in the second row of the signal table
and a resonant second order low-pass filter in the third row of the table.
We clearly see that the signals in the right column
have considerably smaller amplitude than those in the left column.

Why do the amplitudes of filtered noise depend on the sampling rate?
An intuitive answer can be found in the frequency spectra
that are depicted in the bottom row of the table:
Since the frequency spectrum of the high rate signal
covers a larger frequency range,
the energy of the high rate noise is spread over a larger frequency interval.

In order to verify,
that the problem is actually the noise and not the filtering,
we have inserted a centre column,
where all signal processes are performed on a low rate noise,
that was converted to a higher rate
by simply replicating all sample values of the low rate noise four times.
The audio impression is the same as for the sounds in the left column.
The important difference between the upsampled noise in the centre column
and the high rate noise in the right column is,
that the high rate noise consists of \keyword{independent} random values
whereas the random values in the upsampled noise
are \keyword{equal} within blocks of four values.

\begin{figure*}

\newcommand\plotsignal[1]{
% \hspace{-2ex}
\begin{tabular}[c]{r}
\begin{tikzpicture}
\draw[->] (-0.2,0) -- (4.2,0) node[right] {$t$};
\draw[->] (0,-1.8) -- (0,1.8) node[above] {$x(t)$};
\draw[xscale=80,yscale=1.5] plot file {program/#1};
\end{tikzpicture}
\end{tabular}
}

\newcommand\plotspectrum[1]{
% \hspace{-2ex}
\begin{tabular}[c]{r}
\begin{tikzpicture}
\draw[->] (-0.2,0) -- (4.2,0) node[right] {$f$};
\draw[->] (0,-1.8) -- (0,1.8) node[above] {$\hat{x}(f)$};
\draw[xscale=0.18,yscale=600] plot file {program/#1};
\end{tikzpicture}
\end{tabular}
}

\begin{tabular}{ccc}
$x = \render(\hz{11025}, \noise)$ &
$
\begin{array}{rcl}
x &=& \resample (\hz{44100}, \\
&&\quad\render(\hz{11025}, \noise))
\end{array}
$
&
$x = \render(\hz{44100}, \noise)$
\\
% \multicolumn{3}{c}{\code{x}} \\
\plotsignal{noise-low.csv} &
\plotsignal{noise-upsmp.csv} &
\plotsignal{noise-high.csv}
\\
% \multicolumn{3}{c}{movingAverage x} \\
\multicolumn{3}{c}{$\mathrm{firstOrderLowpass}(x)$} \\
\plotsignal{noise-low-1st-order.csv} &
\plotsignal{noise-upsmp-1st-order.csv} &
\plotsignal{noise-high-1st-order.csv}
\\
\multicolumn{3}{c}{$\mathrm{resonantLowpass}(x)$} \\
\plotsignal{noise-low-uni.csv} &
\plotsignal{noise-upsmp-uni.csv} &
\plotsignal{noise-high-uni.csv}
\\
\multicolumn{3}{c}{$\mathrm{absoluteSpectrum}(x)$} \\
\plotspectrum{noise-low-spectrum.csv} &
\plotspectrum{noise-upsmp-spectrum.csv} &
\plotspectrum{noise-high-spectrum.csv}
\end{tabular}
\figcaption{noise-filter}{
Table of three signals and the result of various transformations applied to them.
The three initial signals are
noise at sampling rate \hz{11025},
noise at \hz{11025} upsampled to \hz{44100}
by constant interpolation,
noise at \hz{44100}.
The duration of the sounds is 50~\unit{ms}.
The initial signals are depicted in the first row of the table.
The row below contains the results of applying a first order low-pass filter
with cut-off frequency of \hz{500}.
The third row contains the results of a resonant low-pass filter
from a state-variable filter
with resonance frequency \hz{500}.
The last row contains frequency spectra of the initial sounds.
}
\end{figure*}

\subsection{Contributions}

With our paper we want to contribute the following aspects
for resolving the sampling rate dependence of white noise:
\begin{itemize}
\item Develop a criterion for judging whether
a signal generator or modifier performs similarly in different sampling rates
in \secref{comparability}.
\item Explore some ways of adapting noise to the sampling rate
in \secref{samplerate}.
\item Consider several signal modifiers like
frequency filters, quantisers, click generators
and how they can be made aware of noise input and sampling rates
in \secref{processes}.
\item Discuss in \secref{parameter} by what parameters
a sampling-rate-aware noise generator should be controlled.
\item Give a small guide on choosing a random distribution
in \secref{distribution}.
\end{itemize}

\section{Main Work}

\subsection{Comparability across sampling rates}
\seclabel{comparability}

In natural sounds there is no such thing
as a time quantisation and a sampling rate.
Thus natural signals are commonly modelled by real functions.
But when it comes to signal processing in a digital computer
we need time (and value) discretisation.
Nonetheless we do not want to think about discretisation
when designing a signal processing algorithm.
We like to pretend that there is no sampling
and thus an algorithm without a reference to sampling
can be used both for analogue synthesis
and for digital synthesis at any sampling rate.
\begin{definition}[Abstract signal processing algorithm]
\dfnlabel{abstract-algorithm}
We like to call a signal processing algorithm \keyword{abstract},
if it does not contain any reference to discretisation or a sampling rate.
All quantities in such an algorithm shall be physically meaningful,
e.g.\ time values must be given in seconds
but not as numbers of sampling periods.
An example is the Csound algorithm in \figref{panpipe-csound}.
\end{definition}

Since an abstract signal processing algorithm neglects sampling,
we can use it to describe real functions.
Interpreting an abstract signal processing algorithm
at a given sampling rate means,
that we approximate these real functions by discrete signals.
E.g.\ if we describe a frequency filter as the solution of a differential equation,
then this is an abstract algorithm.
In the digital computer we compute a corresponding difference equation
and this is the interpretation of the abstract algorithm
for a given sampling rate.
We measure the quality of the difference equation solver
by its closeness to the solution of the according differential equation.

For investigation of noise,
real functions are not of much use as a model,
since real functions with stochastic values
are neither continuous nor integrable.
In contrast to that,
there is no problem in computing differences or sums in discrete noise.
We may be able to model noise using stochastic processes,
stochastic differential equations
and generalised measurements of the degree of approximation
between a discrete signal and stochastic function.
But we think that the following approach is easier:

We accept the lack of a discretisation-free model
that we can adapt our discrete computations to.
Instead we ask for comparable results,
when interpreting the same abstract signal processing algorithm
for different sampling rates.
That is, increasing the sampling rate for rendering
shall improve the audio quality
but it shall not alter the timbre of the sound signal.

How can we check, whether a particular discrete interpretation
of an abstract signal processor
generates comparable results for different sampling rates?
We have to convert between the sampling rates.
We cannot add information by upsampling a signal from a low sampling rate,
but we can discard information by downsampling a signal from a high to a low sampling rate.
Downsampling should act as a projection:
It shall maintain the information, that can be represented at the lower rate
and it shall discard the remaining information.

In order to write the comparability requirement a bit more formally,
we like to define $\render(r,A)$,
that denotes a discrete signal,
that is computed from the abstract algorithm~$A$ at sampling rate~$r$.
Think of $A$ being the Csound orchestra definition in \figref{panpipe-csound},
$r$ being the number we pass to the \verb|--sample-rate| option
and $\render$ as being the \code{csound} command.
The sampling rate~$r$ becomes part of the generated signal,
such as it becomes part of the \code{WAVE} file generated by \code{csound}.
Further on we like to denote the resampling of a signal~$x$
from its associated sampling rate to another sampling rate~$r$ by $\resample(r,x)$.
Now we can state:
\begin{criterion}[comparability across sampling rates]
\crtlabel{comparability}
The discrete interpretation (expressed by $\render$)
of an abstract signal processing algorithm~$A$
is called \keyword{comparable across sampling rates} if
\begin{multline*}
\forall r_0 \forall r_1 \quad r_0 \le r_1 \implies \\
   \render(r_0, A) \approx \resample(r_0, \render(r_1, A))
\end{multline*}
\end{criterion}
If $\render(r_0, A)$ computes a band-limited version of $\render(r_1, A)$
and $\resample$ performs perfect resampling,
then ``$\approx$'' could be replaced by ``=''.
However most actual implementations of discrete signal processing
only approximate this ideal world.
For noise it is even worse, since we can hardly create
``the same'' noise at different sampling rates.
Thus we have to interpret ``$\approx$'' even weaker
as an equivalence of some stochastic characteristics.
Although the above criterion is in no way mathematically precise,
it turns out to be a very useful guide for design decisions
in the following sections.

\subsection{Adapt noise to sampling rate}
\seclabel{samplerate}

A simple way to provide noise that behaves similar across different sampling rates,
is to upsample noise from a low rate.
Say, we are satisfied with the range of frequencies
contained in discrete noise at \hz{11025} sampling rate.
For sounds at \hz{44100} we can just upsample that noise from \hz{11025} to \hz{44100}.
This approach trivially generates comparable noise signals
for all sampling rates above \hz{11025},
even with ``$\approx$'' replaced by ``$=$'' in \crtref{comparability}
when we use pseudo-random numbers with the same seed for all sampling rates.
But there are two disadvantages:
\begin{itemize}
\item This way we cannot generate comparable noise for rates below \hz{11025}.
\item We want to increase rendering quality by increasing the sampling rate.
For noise, we like to read ``higher quality''
to mean a larger range of random frequencies.
However, with the upsampling approach we do not automatically get
higher frequency portions in the noise,
when we switch to higher sampling rates.
\end{itemize}

If we generate pseudo-random numbers at the target sampling rate,
then we automatically fill the entire available frequency space.
However, as we have seen in the introduction,
we have to somehow adapt the noise amplitude in order to provide
equal frequency amplitudes.
We still have to live with the drawback,
that this kind of sampling-rate-aware noise
is comparable across sampling rates
only with respect to stochastic parameters
but not in terms of actual approximations.

In the following sections
we will derive the necessary amplitude adjustment
and we will see how other signal processes must be adapted
in order to work nicely with noise.

\subsection{How to further process noise}
\seclabel{processes}

\subsubsection{Frequency Filter}
\seclabel{filtered-noise}

An important way of modifying white noise is frequency filtering.
In analogy to electromagnetic oscillations of light,
filtered noise is called coloured noise.
Pink noise, i.e.\ low-pass filtered noise, can be used as control curve.
White noise filtered by resonant low-pass filters
can produce sounds of wind, echo sounding, or fricatives.

We want to investigate how to adapt noise to sampling rates
such that it behaves similar with respect to frequency filters.
That is, according to \crtref{comparability}
we want to achieve
\begin{multline}
\forall r_0 \forall r_1 \quad r_0 \le r_1 \implies \\
   \render(r_0, \filter(\noise)) \\
   \approx
   \resample(r_0, \render(r_1, \filter(\noise)))
\quad.
\eqnlabel{comparable-filter}
\end{multline}
\begin{comment}
\begin{multline*}
   \spectrum(\render(r_0, \filter(\noise))) \\
   \approx
   \spectrum(\resample(r_0, \render(r_1, \filter(\noise))))
\end{multline*}
\begin{multline*}
   \spectrum(\render(r_0, \filter(\noise))) \\
   \approx
   \clip(r_0, \spectrum(\render(r_1, \filter(\noise))))
\end{multline*}
\begin{multline*}
   \render(r_0, \spectrum(\filter(\noise))) \\
   \approx
   \clip(r_0, \render(r_1, \spectrum(\filter(\noise))))
\end{multline*}
\begin{multline*}
   \render(r_0, \weight(\spectrum(\noise))) \\
   \approx
   \clip(r_0, \render(r_1, \weight(\spectrum(\noise))))
\end{multline*}
\end{comment}

Let us start with the simple example of a moving average filter,
where the arithmetic mean of $w$ successive values is computed.
We model white noise as a sequence of random variables,
that all have the expectation value 0 and the same variance.
The technical term for such a sequence is \keyword{discrete stochastic process}.
The expectation value corresponds to the direct current offset,
whereas the standard deviation (root of the variance)
is the measure of the noise volume.
We start with filtering white noise $X_\text{low}$
at sampling rate $\hz{11025}$.
\begin{IEEEeqnarray*}{rCl}
\separateeqn{\render(\hz{11025}, \filter(\noise)):}
Y_{\text{low},k} &=&
  \frac{1}{w} \cdot \sum_{j=k}^{k+w-1} X_{\text{low},j}
\end{IEEEeqnarray*}
In order to perform the same filter
at the higher sampling rate $\hz{44100}$
and an according white noise $X_\text{high}$,
we have to increase the number of averaged values to $4w$.
\begin{IEEEeqnarray*}{rCl}
\separateeqn{\render(\hz{44100}, \filter(\noise)):}
Y_{\text{high},k} &=&
  \frac{1}{4w} \cdot \sum_{j=k}^{k+4w-1} X_{\text{high},j}
\end{IEEEeqnarray*}
We observe that
\begin{IEEEeqnarray*}{rCl}
\stddev(Y_{\text{low},0})  &=& \frac{1}{w}\cdot\stddev(X_{\text{low},0}) \\
\stddev(Y_{\text{high},0}) &=& \frac{1}{2\cdot w}\cdot\stddev(X_{\text{high},0})
\end{IEEEeqnarray*}
that is, for equal standard deviations of the white noises
the standard deviations of the filtered noises are not equal.
From $\stddev(X_{\text{low},0}) = \stddev(X_{\text{high},0})$
it follows $\stddev(Y_{\text{low},0}) = 2 \cdot \stddev(Y_{\text{high},0})$.

How to resolve this inconsistency?
For the filtered noise the resample operation in \eqnref{comparable-filter}
is essentially a matter of keeping every fourth value.
That is we can require $Y_{\text{low},k} \approx Y_{\text{high},4k}$.
We like to read this as
\[
\forall k\quad
\stddev(Y_{\text{low},k}) = \stddev(Y_{\text{high},4k})
\]
To achieve this, we have to set the standard deviation of $X_k$
proportional to the square root of the sampling rate.
Note, that downsampling of white noise
cannot be done simply by picking values at a coarser grid,
since this skips the necessary limitation of the frequency band.
\[
\Var(X_k) \sim r
\]

Now we move on to general frequency filters.
They become most simple in the frequency spectrum
(just a weighting of the spectral values)
and also downsampling is only matter of shortening the spectrum.
Thus we like to interpret ``$\approx$'' in \eqnref{comparable-filter}
as comparing the frequency spectra.

Let $X$ be a sequence of $n$ random variables,
that all have the expectation value 0 and the variance $y^2$.
The noise sampling rate is $r$ and
it may have a physical unit such as \unit{Hz}.
The discrete frequency spectrum $\IDFT(X)$ is defined by
\[
\IDFT(X)_k =
\frac{1}{r}\cdot\sum_{j=0}^{n-1}
   X_j\cdot\exp\left(2\pi i\cdot \frac{j\cdot k}{n}\right)
\quad.
\]
We have to interpret ``$\approx$'' in \eqnref{comparable-filter}
as the equality of the standard deviations of the \Fourier{} coefficients,
because we cannot expect similarity of observed frequency amplitudes.
Since the random variables in $X$ are independent,
their variance is additive.
\begin{IEEEeqnarray*}{rCl}
\Var(\IDFT(X)_k)
 &=& \frac{1}{r^2}\cdot\sum_{j=0}^{n-1} \Var(X_j) \\
 &=& n\cdot\frac{y^2}{r^2}
\eqnlabel{dft-noise} \\
\stddev(\IDFT(X)_k) &=& \sqrt{n} \cdot \frac{y}{r}
\end{IEEEeqnarray*}
Since $n$ depends on the sampling-rate via $n=l\cdot r$,
and we must compare signals of the same length~$l$,
we have to substitute $n$.
\begin{IEEEeqnarray*}{rCl}
\stddev(\IDFT(X)_k) &=& \sqrt{\frac{l}{r}} \cdot y
\end{IEEEeqnarray*}
That is, if noise of duration~$l$ at sampling rate~$r$
shall have spectral values with standard deviation~$c$
(i.e.\ $c = \stddev(\IDFT(X)_k)$),
then we have to choose
\begin{equation}
y = \sqrt{\frac{r}{l}}\cdot c
\quad.
\eqnlabel{rate-dependent-amplitude-dft}
\end{equation}
The amplitude of the noise is proportional
to the square root of the sampling rate.

The part in \eqnref{rate-dependent-amplitude-dft}
that does not depend on the sampling rate is $\frac{c}{\sqrt{l}}$.
We like to call that the \keyword{noise voltage spectral density value}.
Usually spectral density is a function defined for real signals.
In the following definitions we want to adapt
the required terms from real signals to discrete ones.
\begin{definition}[Wide-sense stationary discrete stochastic process]
A discrete stochastic (or random) process~$X$,
where all elements have expected value~$0$
is called \keyword{stationary in a wide sense}
if the covariance between its elements
depends only on the distance but not on the time point.

Expressed in formulas:
\begin{IEEEeqnarray*}{r'rCl}
\forall k & \expectedvalue(X_k) &=& 0 \\
\forall k\forall d
 & \expectedvalue(X_0 \cdot X_d) &=& \expectedvalue(X_k \cdot X_{k+d})
\end{IEEEeqnarray*}
\end{definition}
The white noise signals, that we consider in this paper,
and also filtered white noise signals
are always discrete stochastic processes in a wide sense.
\begin{definition}[Autocovariance function]
For a wide-sense discrete stochastic process~$X$
we define the \keyword{autocovariance} function $R_X$
(often called \keyword{autocorrelation})
as the covariances between signal values depending on their distance.
\begin{IEEEeqnarray*}{rCl}
R_X(d) &=& \expectedvalue(X_0 \cdot X_d)
\end{IEEEeqnarray*}
\end{definition}
This captures all possible values of covariances between signal values,
because the wide-sense stationarity warrants time-invariance of the covariances.
\begin{definition}[Noise power spectral density]
The noise spectral density of a wide-sense discrete stochastic process~$X$
is the spectrum of the autocovariance function of $X$.
\begin{equation*}
\NSD(X) = \IDFT(R_X)
\end{equation*}
\end{definition}
Since the signal values of white noise are independent,
the autocovariance function is an impulse at time point zero
with height $\Var(X_0)$.
Its spectrum is a constant function with value $\frac{\Var(X_0)}{r}$.
According to \eqnref{dft-noise} that is equal to $\frac{c^2}{l}$.
We like to call this value
the \keyword{noise power spectral density value of white noise}.
The voltage spectral density is the square root of the power spectral density.
We want to use this as the parameter,
that describes the amplitude of white noise in a sampling-rate-aware way,
and thus give it a symbol, namely $\VSD$.
\begin{equation}
\VSD = \frac{\stddev(X_0)}{\sqrt{r}}
\eqnlabel{rate-dependent-amplitude}
\end{equation}

\subsubsection{Quantisation}

Quantising noise in time direction
is a way, to give noise a pitch characteristic,
when using small quantisation periods,
and is useful as control curve
for large quantisation periods.
% random melodies, random amplitudes, random locations in space, random filter frequencies

For reasons of simplicity we will consider
quantisation with fixed quantisation periods
that are integral multiples of the sampling period.
For a discrete input signal~$x$ with sampling rate~$r$
and quantisation period~$t$,
and $d$ being the quantisation period measured in units of the sampling period,
that is $d = t\cdot r, d\in\N$,
we could simply define
\begin{equation}
\quantise(x)_k = x_{k - (k \bmod d)}
\quad.
\eqnlabel{naive-quantisation}
\end{equation}
This would yield a constant amplitude of $\quantise(x)$
for constant quantisation period~$t$ and varying sampling rate~$r$
if the amplitude (standard deviation) of $x$ would not depend on $r$.
However if we quantise sampling-rate-aware noise
as described in \secref{filtered-noise} this way,
then the amplitude of the quantised noise
with respect to a constant quantisation period
will increase proportional to the square root of the sampling rate.
In this respect an amplitude that increases with the sampling rate is not good,
since the quantisation period acts like an artificial low sampling rate
represented at a high sampling rate,
and this quantisation period does not depend on the actual sampling rate.

We can avoid a growing amplitude by averaging over the quantisation period.
\begin{comment}
\[
\quantise(x)_k =
   \sum_{j=k - (k \bmod d)}^{k+d-1 - (k \bmod d)} x_j
\]
\end{comment}
\begin{IEEEeqnarray*}{rCl}
\quantise(x)_{\kappa} &=& q_{\lfloor \kappa/d \rfloor}
   \eqnlabel{averaging-quantisation} \\
\text{with }q_k
   &=& \frac{1}{d}\cdot\sum_{j=k\cdot d}^{(k+1)\cdot d - 1} x_j
\end{IEEEeqnarray*}
In the following proof we show that the amplitude
of quantised sampling-rate-aware white noise~$X$
% (a sequence of random variables)
with spectral density as in \eqnref{rate-dependent-amplitude}
does not depend on the sampling rate.
That is, we check the criterion in \crtref{comparability}
where we interpret ``$\approx$'' as comparing the standard deviation
(= the amplitude) of the quantised noise.
\begin{IEEEeqnarray*}{rCl}
Q_k &=& \frac{1}{d}\cdot\sum_{j=k\cdot d}^{(k+1)\cdot d - 1} X_j \\
% \stddev\left(\quantise(x)_k\right)
\Var\left(Q_k\right)
 &=& \frac{d}{d^2} \cdot \Var(X_{k\cdot d})
\ = \frac{1}{t\cdot r} \cdot r\cdot \VSD^2 \\
\stddev\left(Q_k\right)
 &=& \frac{\VSD}{\sqrt{t}}
\end{IEEEeqnarray*}
We see that the amplitude of the quantised noise
grows proportionally to the square root
of the quantisation frequency~$\frac{1}{t}$.
This is compliant with our sampling-rate-aware white noise generation,
where we want that noise with more frequency content is also louder.
In fact quantisation can be seen as downsampling,
that includes an appropriate low-pass filter,
with subsequent upsampling by constant interpolation.

A disadvantage of averaging quantisation as in \eqnref{averaging-quantisation}
is that in real-time processing
it delays the signal by one quantisation period~$t$,
whereas the simple quantisation is in \eqnref{naive-quantisation}
does not cause such a delay.

\subsubsection{Random clicks (impulse noise)}
\seclabel{random-clicks}

There is another important kind of sounds that is based on randomness:
Randomly occurring impulses.
By subsequent processes like frequency filters
we can change the characteristic to several natural sounds.
Examples are the sound of raindrops, hail,
or the \person{Geiger}-\person{M\"uller}-counter
for measurement of ionising radiation.

A simple approach to generate random impulses from white noise is as follows:
From white noise at sampling rate~$r$
with samples that are uniformly distributed between $-y$ and $y$
we want to obtain random impulses with a frequency~$f$.
We generate an impulse in the output signal
whenever the white noise sample is
in the interval $\left[-\frac{yf}{r}, \frac{yf}{r}\right]$.

This approach is very simple but it has several drawbacks:
\begin{itemize}
\item It is bound to input noise with uniformly distributed sample values.
\item The threshold value $\frac{yf}{r}$ for given frequency~$f$
depends on the sampling rate~$r$.
If the input white noise is sampling-rate-aware and uniformly distributed,
then its sample values cover $[-k\sqrt{r}, k\sqrt{r}]$
(i.e.\ $y = k\sqrt{r}$)
for sampling-rate independent $k$
and the threshold must be $\frac{kf}{\sqrt{r}}$.
\item For smooth input signals (no noise) we get clusters of impulses,
what in discrete signal processing means,
that we get signals consisting of constant pieces
rather than separated impulses.
\item We can only control the overall frequency of impulses
but not the degree of randomness.
\end{itemize}

All of these problems can be solved using $\Delta\Sigma$-modulation
as in \figref{delta-sigma-modulation}.
We integrate white noise with a positive direct current offset
until it exceeds a threshold.
At this time point we emit an impulse in the output signal
and then start integrating with cleared accumulator again.
We repeat this procedure in an endless loop.

In the discrete implementation of the $\Delta\Sigma$-converter
the integrator \framebox{$\int$} is a cumulative sum,
the comparator \framebox{$>y$}
emits an impulse with a height related to $y$,
if the input exceeds the threshold~$y$ and zero otherwise,
the \framebox{delay} delays by one sampling period
in order to make the feedback possible,
and the \textcircled{$-$} subtracts
the fed-back impulse signal from the input,
such that the integrator is reset after every emitted impulse.

% recursive equation

\tikzstyle{process}=[inner sep=1.5ex,rectangle,draw=black!50,fill=black!3,thick]
\tikzstyle{dot}=[inner sep=1pt,circle,fill=black]
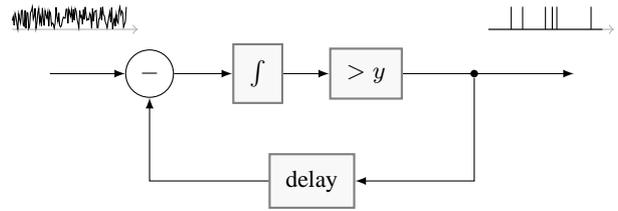
\begin{figure}
\begin{center}
\begin{tikzpicture}[node distance=10ex]
  \node[]            (output)      {};
  \node[dot]         (branch)      [left of=output]     {};
  \node[process]     (compare)     [left of=branch]     {$>y$};
  \node[process]     (integrate)   [left of=compare]    {$\int$};
  \node[circle,draw] (difference)  [left of=integrate]  {$-$};
  \node[]            (input)       [left of=difference] {};
  \node[process]     (delay)       at ([xshift=-5ex,yshift=-10ex] compare) {delay};
\pgfsetarrowsend{latex}
  \draw (input)           -- node[above=1.5em,pos=-0.5](input-wave){}
                             (difference.west);
  \draw (difference.east) -- (integrate.west);
  \draw (integrate.east)  -- (compare.west);
  \draw (compare.east)    -- node[above=1.5em,midway](output-wave){}
                             (output);
  \draw (delay.west)      -| (difference.south);
  \draw (branch)          |- (delay.east);

\pgfsetarrowsend{}

% mapM_ print $ zip (map (\r -> realToFrac r :: Float) [0,0.04..5::Rational]) (Random.randomRs (0,1::Float) (Random.mkStdGen 32))
  \begin{scope}[shift={(input-wave)},scale=0.3]
  \draw[->,black!30] (0,0) -- (5.5,0);
  \draw plot coordinates{
      (0.0,0.39061373)
      (4.0e-2,0.5491117)
      (8.0e-2,0.18063757)
      (0.12,0.6977923)
      (0.16,0.5431789)
      (0.2,0.24685007)
      (0.24,2.6195168e-2)
      (0.28,0.68086505)
      (0.32,0.44515383)
      (0.36,0.27486205)
      (0.4,0.8329313)
      (0.44,0.8408313)
      (0.48,0.1937409)
      (0.52,0.1461975)
      (0.56,0.8314311)
      (0.6,5.1748455e-2)
      (0.64,0.74603206)
      (0.68,0.72929573)
      (0.72,4.5295984e-2)
      (0.76,0.1823709)
      (0.8,0.7632302)
      (0.84,0.8195398)
      (0.88,0.98861945)
      (0.92,0.44240162)
      (0.96,0.5922496)
      (1.0,0.9318246)
      (1.04,5.558309e-2)
      (1.08,0.27662194)
      (1.12,0.72695315)
      (1.16,0.9212217)
      (1.2,0.30458164)
      (1.24,0.5570083)
      (1.28,0.13947546)
      (1.32,0.93885696)
      (1.36,0.46760225)
      (1.4,0.16948941)
      (1.44,0.81575966)
      (1.48,0.64150214)
      (1.52,0.49859917)
      (1.56,0.7641877)
      (1.6,0.14129218)
      (1.64,0.7977605)
      (1.68,0.684986)
      (1.72,0.18602917)
      (1.76,0.27862352)
      (1.8,0.23791155)
      (1.84,5.5978805e-2)
      (1.88,0.26019633)
      (1.92,0.97633064)
      (1.96,0.55528784)
      (2.0,0.34306526)
      (2.04,0.7276543)
      (2.08,0.16213647)
      (2.12,0.10440126)
      (2.16,0.18030006)
      (2.2,0.40402338)
      (2.24,0.5380113)
      (2.28,0.5030661)
      (2.32,0.49254087)
      (2.36,0.8859041)
      (2.4,0.89344454)
      (2.44,0.26869938)
      (2.48,0.8848692)
      (2.52,0.72334564)
      (2.56,0.56806856)
      (2.6,0.8591485)
      (2.64,0.31173456)
      (2.68,0.8883259)
      (2.72,0.40846375)
      (2.76,0.9101517)
      (2.8,0.80354387)
      (2.84,0.46834472)
      (2.88,0.27541476)
      (2.92,7.630244e-2)
      (2.96,0.8282986)
      (3.0,0.53946644)
      (3.04,0.6063649)
      (3.08,0.22144791)
      (3.12,0.8657899)
      (3.16,0.44108614)
      (3.2,0.986755)
      (3.24,0.62553465)
      (3.28,0.4058249)
      (3.32,0.47048932)
      (3.36,0.43584383)
      (3.4,0.7105382)
      (3.44,0.6172842)
      (3.48,4.3167293e-2)
      (3.52,0.92599714)
      (3.56,0.36051494)
      (3.6,0.7580808)
      (3.64,0.81017995)
      (3.68,0.8367493)
      (3.72,7.559949e-2)
      (3.76,7.4644476e-2)
      (3.8,0.4741462)
      (3.84,0.4864546)
      (3.88,0.9616835)
      (3.92,0.4205309)
      (3.96,0.98898613)
      (4.0,0.6446306)
      (4.04,0.59457505)
      (4.08,0.33410865)
      (4.12,0.23318917)
      (4.16,7.5772434e-2)
      (4.2,0.34355736)
      (4.24,0.57338595)
      (4.28,0.7742748)
      (4.32,0.3233459)
      (4.36,0.9646327)
      (4.4,2.1863878e-2)
      (4.44,0.28144285)
      (4.48,0.4714442)
      (4.52,0.14407739)
      (4.56,0.6666917)
      (4.6,0.35866398)
      (4.64,0.27224272)
      (4.68,0.8275545)
      (4.72,0.9916374)
      (4.76,0.64898145)
      (4.8,0.8106249)
      (4.84,0.7330942)
      (4.88,0.1068064)
      (4.92,0.22920701)
      (4.96,0.34037143)
      (5.0,0.99322593)
    };
  \end{scope}

  \begin{scope}[shift={(output-wave)},scale=0.3]
  \draw[->,black!30] (0,0) -- (5.5,0);
  \draw plot coordinates{(0,0) (5,0)};
  \draw plot coordinates{(1.0,0) (1.0,1)};
  \draw plot coordinates{(1.5,0) (1.5,1)};
  \draw plot coordinates{(2.5,0) (2.5,1)};
  \draw plot coordinates{(2.8,0) (2.8,1)};
  \draw plot coordinates{(3.0,0) (3.0,1)};
  \draw plot coordinates{(4.5,0) (4.5,1)};
  \end{scope}
\end{tikzpicture}
\end{center}
\figcaption{delta-sigma-modulation}{%
$\Delta\Sigma$-modulation.
}
\end{figure}
The expectation value of the input white noise,
i.e.\ the direct current offset,
determines the frequency of peaks in the output,
whereas the variance of the noise
determines the degree of randomness of peak distribution.

We want to prove, that impulse generation from white noise
via $\Delta\Sigma$-modulation
yields comparable frequency and randomness of impulses across sampling rates.
The noise generation and the integration
are the only operations that adapt to the sampling rate.
That is, it suffices to show that the integral over a fixed duration~$t$
of a sequence~$X$ of identically distributed random variables
with standard deviations as in \eqnref{rate-dependent-amplitude}
has an expected value and a variance
that does not depend on the sampling rate~$r$.
For simplicity $t\cdot r$ shall be an integer.
\begin{IEEEeqnarray*}{rCl}
\int_0^t X
  &=& \frac{1}{r}\cdot \sum_{k=0}^{t\cdot r - 1} X_k \\
\expectedvalue\left(\int_0^t X\right)
  &=& t\cdot \expectedvalue(X_k) \\
%  &=& \frac{t\cdot r}{r} \expectedvalue(X_k) \\
\Var\left(\int_0^t X\right)
  &=& \frac{t\cdot r}{r^2}\cdot r\cdot \VSD^2 \\
\stddev\left(\int_0^t X\right)
  &=& \sqrt{t}\cdot \VSD
\end{IEEEeqnarray*}

We have still not answered the question,
what kind of impulses the $\Delta\Sigma$-modulator shall generate.
If we use impulses with one sampling period as duration,
then the height of the impulses must be chosen,
such that when fed back it resets the accumulator in the integrator.
To this end let us consider the involved physical units:
Let the input signal have time unit~\unit{s} and amplitude unit~\unit{V}.
Then the integrated signal has amplitude unit~\unit{Vs}
and so the threshold in the comparator must have this unit, too.
Thus the impulse, that the comparator generates,
must have an area equal to the threshold~$y$,
in order to clear the accumulator.
Since its width is $\frac{1}{r}$, its height must be $r\cdot y$.

This way the impulses have sizes
such that they represent the area of the input signal
over the pauses between the impulses.
This means, that smoothing the impulse train
yields a signal similar to the input signal after smoothing.
This property is actually the key
for using $\Delta\Sigma$-modulator in digital-analogue converters.

An alternative approach for generating random impulses
is to choose the pauses between the impulses
according to pseudo-random numbers % that are normally distributed
with expectation value ``average silence duration between impulses''
and variance ``degree of randomness''.
That is, strictly spoken
it is not necessary to generate random impulses from white noise.
\begin{comment}
However, in a function library that provides white noise,
the user can try to generate random impulses from white noise himself.
We want to make sure that such an algorithm
does not dependent on a particular sampling rate by accident.
By providing sampling-rate-aware white noise,
$\Delta\Sigma$-modulation
and sample-wise comparison without access to the sampling rate,
we are relatively safe in this respect.
Really?
No, I think, a non-linear process like a comparator remains dangerous.
\end{comment}
However, on the one hand we wanted to show,
that random impulse generation from sampling-rate-aware white noise
is possible in a way, that is itself sampling-rate-aware.
On the other hand we wanted to point out
that the use of a comparator (as in the beginning of this section)
can lead to signal algorithms,
that depend on the sampling rate by accident.

% \subsubsection{Zero detection}

\subsection{Noise parameters}
\seclabel{parameter}

% We need a parameter that describes the right thing
% independent from sampling rate and signal length.

In principle a sampling-rate-aware white noise generator
with amplitude unit $\unit{V}$ and time unit $\unit{s}$
must be controlled by a parameter
with unit $\unit{V}/\sqrt{\unit{Hz}}$ or $\unit{V}\cdot\sqrt{\unit{s}}$,
that we called \keyword{voltage spectral density value} $\VSD$
(see \eqnref{rate-dependent-amplitude}).
However that is both unintuitive and
unsupported by the usual implementations
of physical dimensions in programming languages
(\cite{syme2007fsharp,buckwalter2010dimensional-0.8.0.1})
where exponents of units must be integers.
It is unintuitive, because it is not simple to choose a number
that yields a reasonable amplitude.
E.g.\ we must choose $\VSD = \frac{1~\unit{V}}{\sqrt{\hz{44100}}}
\approx 4.67~\unit{mV}\cdot\sqrt{\unit{s}}$
in order to get standard deviation $1~\unit{V}$
when rendering at sampling rate \hz{44100}.
We avoid the factional powers in units using the squared parameter,
that is the \keyword{power spectral density value}
with unit $\unit{V}^2\cdot\unit{s}$.
This is even less intuitive, since doubling the noise amplitude
means using four times of the power density value.
In our experience a very intuitive solution is
to use two parameters $y$ and $f$ with the units $\unit{V}$ and $\unit{Hz}$.
They mean that at sampling rate~$f$ the variance shall be $y$
and the variance for other sampling rates shall be adjusted accordingly.
Given these parameters the noise generator must compute samples
of random variables~$X$
with \[
\expectedvalue(X) = 0\qquad
\stddev(X) = y\cdot\sqrt{\frac{r}{f}}
\quad.\]

% http://www.ecircuitcenter.com/Circuits/Noise/Noise_Analysis/res_noise.htm
%   Noise Power Spectral Density
%   Noise Voltage Spectral Density

\subsection{Random distribution}
\seclabel{distribution}

So far we did not need to consider particular random distributions,
because we only needed additivity of the variance of random variables.
The choice of the random distribution
does not have an effect on the shape of the frequency spectrum.
If the random variables of a noise signal are independent from each other,
then all spectral values have the same variance.
However, since the human ear performs something
more like a short-time \Fourier{} transform,
a random distribution considerably different from normal distribution
may generate single clicks, that can be heard.

Because in signal processing many operations
like frequency filtering, integration, mixing
involve addition, it is likely that the Central Limit Theorem applies.
The result of applying signal algorithms to white noise
are likely to yield random variables with random distributions
close to normal distribution.
For reasons of consistency we may thus prefer normal distributions from the beginning.
A very simple way to approximate
normally distributed random variables with variance 1
is to add three uniformly $[-1,1]$-distributed random variables.
The actual distribution has the shape of a quadratic B-spline.
If speed matters,
then white noise with uniformly distributed random variables is the best choice.
This is what pseudo-random number generators create.

\section{Related Work}

A wide range of the literature considers noise
that arises as an undesirable artefact of signal processing.
This part of the literature identifies properties
that allows to compare the behaviour of electronic circuits with respect to noise
and to separate noise and non-noise portions of a signal.
In this literature the notion of the noise spectral density is well-known.
\cite[Chapter 2]{kester2005dataconversion}
% \cite[Section 2.3]{kester2005dataconversion}

% (except investigations on random distributions)
Intended generation of noise is not equally popular.
A notable exception is \cite{hanna2002noise},
where the authors construct noise
by mixing sine waves at random frequencies.
Consequently they use a custom definition of noise spectral density,
where the density is the number of sine waves
divided by the width of the frequency band.
By using the same ratio of present frequencies per band across sampling rates
they can create sampling-rate-aware noise in a trivial way.
The sinusoidal model allows to control the noise colour in an intuitive way,
but for white noise, it is computationally more intensive,
even when using a Fast Fourier Transform,
than our approach of just adapting the amplitude of a random-number sequence.

\newcommand\parameter[1]{\textit{#1}}

Although the notion of the noise spectral density is well-known in the literature,
we could not find the conclusion,
that discrete white noise should be generated
with a variance proportional to the sampling rate.
As mentioned in the introduction
it is also not implemented in common software synthesizers.
We have tested
Csound-5.10.1 \cite{vercoe2004csound},
SuperCollider-3.3.1 \cite{mccartney1996supercollider},
ChucK-1.2.0.8 \cite{wang2004chuck}.
None of these packages advertises to be sample-rate-aware,
although the use of physically motivated parameters suggest that they are.
However physical parameters such as time in seconds
and frequency in Hertz
% and amplitude in volts
are mixed with low-level parameters
like plain digital filter parameters,
e.g. Csound:\code{noise}:\parameter{kbeta},
SuperCollider:\code{OnePole}:\parameter{coef},
ChucK:\code{BiQuad}:\parameter{a0}.

\begin{comment}
Csound-5.13 mentions a case of unintended sampling-rate-dependency
for the lorenz opcode.

Csound-5.13 manual does not mention spectral density

Csound \cite{vercoe2004csound} does not seem to be intended
for signal processing algorithms that behave similarly
at different sampling rates and sample value ranges.
Its opcodes usually mix machine-oriented parameters and physically motivated parameters.
E.g.\ the amplitude 1 is not interpreted as the maximum of the sample value range,
as the common sampled sound formats imply.
Instead it is interpreted literally.
This means, that amplitude 1 in Csound
is appropriate for floating point sample formats,
but it is too quiet for 8 bit, 16 bit or 24 bit samples
since it uses only the least significant bits.
Another example is the \code{noise} opcode
that combines a white noise generator with a first order low-pass filter.
However the low-pass filter is not controlled by its cut-off frequency
but by a plain feedback factor.
\end{comment}
In Csound the \code{noise} opcode
with disabled smoothing (parameter \parameter{kbeta}=0) generates white noise.
It does not adapt its amplitude to the sampling rate.
This applies to all other of Csound's white noise generators,
that provide different distributions of the random variables
(opcodes \code{gauss}, \code{unirand}, \code{linrand}, \code{cauchy}, \dots).
It also applies to the white noise generators
in SuperCollider (\code{WhiteNoise}) and ChucK (\code{Noise}).

If we want to get coloured noise,
we can call specialised noise generators in those packages.
E.g. there is the Csound opcode \code{pinkish} in the default mode
and the SuperCollider unit generator \code{PinkNoise},
that generate pink noise following a multi-scale scheme
(\person{Moore}/\person{Voss}-\person{McCartney} method).
Although not strictly comparable across different sampling rates,
because at lower sampling rates there are more low-frequency components,
the generated noise is almost comparable across different sampling rates.
\begin{comment}
pinkish seems respects sampling rate,
although in the C code the sampling rate is not mentioned.
In pitch.c the value \verb|p->grd_Scalar| is initialised with 1/pmax.
The difference between 44100 Hz pinkish noise and 22050 Hz pinkish noise
is white noise at 44100 Hz.
Maybe we cannot hear the difference
between smoothed pinkish noise at 44100 Hz and 22050 Hz,
because that difference is quite small white noise.

see http://www.firstpr.com.au/dsp/pink-noise/
\end{comment}

% verified with grep and google
To our surprise we have not found time quantisation
in Csound, SuperCollider and ChucK.
Thus these packages have no problem
in combining a noise generator with a quantisation.
% randh: quantised noise
% randi: interpolated low-rate noise
% jitter: random segments
However Csound:\code{randh}, SuperCollider:\code{LFNoise0}
and ChucK:\code{SubNoise}
provide quantised noise at a given rate.
By design these produce comparable results across different sampling rates.

% verified with grep and google
We could also not find delta-sigma modulation
in Csound, SuperCollider and ChucK.
Since SuperCollider does not allow short-time feedback,
delta-sigma modulation cannot be build from other components.
Nonetheless generation of random impulses is possible with
Csound:\code{mpulse} with \code{rand} input and
SuperCollider:\code{Dust}.
% and in ChucK with a combination of \code{Noise} and \code{Impulse}.
The frequency of impulses is sample-rate-aware,
% randomness can be controlled by rand properties,
but the area of generated impulses varies across sampling rates.

There are also software synthesizers like Timidity and FluidSynth
that are designed for SoundFonts.
SoundFont is a format for archiving sampled sounds
together with loop points, envelopes and
post-processing features like frequency filters.
Since SoundFont-2 does not seem to support a noise generator,
noise must be provided as a sampled sound.
If resampling is implemented properly (i.e. including band-limitation),
then noise is automatically adapted to the sampling rate.
Of course the range of frequencies contained in noise
can never be larger than the noise contained in the stored sampled sound.

Digital hardware synthesizers have a fixed sampling rate
and thus they do not need to adapt to different sampling rates.
The same applies to analogous synthesizers
that do not have a sampling rate at all.

In \cite{schulze1989musikelektronik}
the authors describe an electronic circuit called ``Noise Manipulator''
that creates random impulses from white noise.
It uses the following signal algorithm:
\[
\mathrm{monoflop}(\mathrm{comparator}(y, \mathrm{pinknoise}))
\]
This means: Pink noise is converted to a rectangular signal
using a comparator with threshold~$y$.
Then the monoflop converts low$\rightarrow$high jumps to impulses.
The frequency and randomness of the impulses
depend on the threshold~$y$ and the spectrum of the pink noise
in a non-obvious way.
In analogue signal processing
there is no infinitesimally short \person{Dirac} impulse
and no sampling period.
That is we must explicitly assign a duration and a height to impulses.
It is certainly worth to also generate impulses
with a definite duration in discrete signal processing.

\begin{comment}
Maybe related:
Tim Stilson and Julius Smith:
Alias-Free Digital Synthesis of Classic Analog Waveforms
\end{comment}

\section{Conclusions and Future Work}

% \subsection{Conclusions}

In our paper we found that it is useful to adapt the amplitude of white noise
proportionally to the square root of the sampling rate
in order to achieve a consistent audio impression
across different sampling rates.
To speak in terms of the sound designer in the introduction
(\secref{introduction}):
He extends the amplitude parameters of all of his noise generators
by the factor $\sqrt{\frac{r}{\hz{44100}}}$
as in \figref{panpipe-adapt-csound}
where $r$ is the sampling rate.
This way the noise amplitudes at \hz{44100}
remain as he found them in the course of developing the signal algorithms.
For other sampling rates the amplitude grows proportionally
to the square root of the sampling rate.
\begin{figure}
Orchestra file \path{panpipe-adapt.orc}
\begin{verbatim}
nchnls = 1

instr 1
iamp = 30000
anoise   noise    0.3*iamp*sqrt(sr/44100), 0
acolored lowpass2 anoise, 440, 10
aosci    oscils   0.5*iamp, 440, 0
out      aosci+acolored
endin
\end{verbatim}
% Score file \path{panpipe-adapt.sco}
% \verbatiminput{csound/panpipe-adapt.sco}
\figcaption{panpipe-adapt-csound}{
Modified version of the Csound program in \figref{panpipe-csound}
that automatically adapts to the sampling rate.
}
\end{figure}

This solves the problem for the sound designer.
However the signal algorithm now contains a reference to the sampling rate.
That is, according to \dfnref{abstract-algorithm} it is no longer abstract.
E.g.\ it would not be possible to translate the algorithm
and its parameters to an analogue synthesizer.
To this end we would need a noise generator
with the noise voltage spectral density as parameter
as in \secref{parameter}.

Additionally we have checked, that further processing steps of the noise
like filtering, quantisation and impulse generation,
can maintain the auditory experience across sampling rates
when implemented properly.

% \subsection{Future work}

As seen in section \secref{random-clicks}
it is still possible to accidentally develop signal algorithms
that produce considerably different results across sampling rates.
Especially non-linear operations like comparators are problematic.
We have to further investigate how to reduce that risk
while remaining able to implement all interesting signal processing algorithms.

\section{Acknowledgments}

Many thanks go to my colleague Alexander Hinneburg for proof reading
and discussing the draft of the paper.

% Holger hat mir das Data-Conversion-Handbook empfohlen

\bibliographystyle{IEEEbib}
\bibliography{audio,haskell,thielemann,hackage,literature}

\end{document}